\pdfoutput=1
\RequirePackage{ifpdf}
\ifpdf 
\documentclass[pdftex,a4paper]{sigma}
\else
\documentclass[a4paper]{sigma}
\fi

\usepackage{mathrsfs}

\begin{document}

\newcommand{\SUM}[3]{{\sideset{}{_{#1}}\sum\limits_{#2}^{#3}}}

\newcommand{\Dtheta}[1][1]{{\displaystyle \theta{\vbox to1.7ex{}}' \nobreak{}_{\mskip-7.5mu#1}}\mskip-1.5mu{\vbox
to1.7ex{}}\relax}

\newcommand{\Dvartheta}[1][1]{{\displaystyle\vartheta{\vbox to1.7ex{}}' \nobreak{}_{\mskip-7.5mu#1}}\mskip-1.5mu{\vbox
to1.7ex{}}\relax}

\def\DEF{\mathrel{\vcenter{\hbox{$:$}}{=}}} \def\FED{\mathrel{{=}\vcenter{\hbox{$:$}}}}

\def\Partial#1#2{\frac{\partial #1}{\partial #2}}

\def\what#1{\widehat{#1} }

\allowdisplaybreaks

\renewcommand{\thefootnote}{$\star$}

\renewcommand{\PaperNumber}{035}

\FirstPageHeading

\ShortArticleName{On a~Quantization of the Classical~$\theta$-Functions}

\ArticleName{On a~Quantization of the Classical $\boldsymbol{\theta}$-Functions\footnote{This paper is a~contribution to
the Special Issue on Algebraic Methods in Dynamical Systems.
The full collection is available at
\href{http://www.emis.de/journals/SIGMA/AMDS2014.html}{http://www.emis.de/journals/SIGMA/AMDS2014.html}}}

\Author{Yurii V.~BREZHNEV}

\AuthorNameForHeading{Yu.V.~Brezhnev}

\Address{Tomsk State University, 36 Lenin Ave., Tomsk 634050, Russia}
\Email{\href{mailto:brezhnev@mail.ru}{brezhnev@mail.ru}}

\ArticleDates{Received January 31, 2015, in f\/inal form April 17, 2015; Published online April 28, 2015}

\Abstract{The Jacobi theta-functions admit a~def\/inition through the autonomous dif\/fe\-rential equations (dynamical
system); not only through the famous Fourier theta-series.
We study this system in the framework of Hamiltonian dynamics and f\/ind corresponding Poisson brackets.
Availability of these ingredients allows us to state the problem of a~cano\-ni\-cal quantization to these equations and
disclose some important problems.
In a~particular case the problem is completely solvable in the sense that spectrum of the Hamiltonian can be found.
The spectrum is continuous, has a~band structure with inf\/inite number of lacunae, and is determined by the Mathieu
equation: the Schr\"odinger equation with a~periodic cos-type potential.}


\Classification{14H70; 33E05; 33E10; 37N20; 37J35; 81S10}

\renewcommand{\thefootnote}{\arabic{footnote}}
\setcounter{footnote}{0}

\medskip

\hfill\parbox{6cm}{\footnotesize\textit{Why do we attach importance to a possibility to
describe the classicallimit through theta-functions? \ldots\\
Thus, when quantizing integrable models, sooner or later
we should come to a quantization of theta-functions}~\cite[\S~1]{smirnov}.}

\section{Introduction}

Presently, the only known def\/inition and introductory motivation to the famous theta-functions are through the
corresponding Fourier series
\begin{gather*}
\theta(z|\tau)= \SUM{k}{-\infty}{\infty} {\mathrm{e}}^{\pi{\mathrm{i}} (k^2\tau+2kz)}.
\end{gather*}
Such series appear in numerous areas of mathematics and physics as building blocks for re\-pre\-senta\-tion of solutions
and also have a~great interest in their own rights because of rich structural and applied properties.
To be precise, there are some mathematical arguments to the series structure which come from considering the group and
functional properties of Abelian varieties.
The displayed Fourier structure then becomes, in a~certain sense, unique and is generalized to multidimensional cases.
In this work we shall deal only with Jacobian, i.e., 1-dimensional theta-functions.

There is one imperfection of these `serial' ways of introducing the~$\theta$-functions however.
These def\/initions are \emph{pure functional}, not invariant, and do not even look like a~def\/inition for any special
function of mathematical physics.
The latter is determined usually by a~certain dif\/ferential equation whereas the~$\theta$-series are not;
e.g., they appear (always via theta-ratios) to represent solutions of many integrable models.
On the other hand, we know that property of the model to be algebraically integrable~\cite{vanhaecke} is an invariant
property (commuting operators) and is of a~radically dif\/ferent kind from those models which are solvable in terms of
various special functions~\cite{brezhnev2}.

A further aspect is the fact that~$\theta$-functions are the nearest generalization of the elementary ones~-- harmonic
sine/cosine-functions~-- which are related to the simplest classical solvable model and extensively used, through the
harmonic quantum prototype
\begin{gather*}
\Psi''-x^2\Psi=E \Psi,
\end{gather*}
in the quantum f\/ield theories.
Their low energy limits and excitations are treated as quantum mechanics with the f\/ield boson particles: statements of
this elementary model.

Yet another facet is a~relation with~$\theta$-series quotients.
It is known that basic~$\theta$-quotients are proportional to Jacobi's elliptic ${\rm sn}$-, ${\rm cn}$-, and
${\rm dn}$-functions~\cite{abr, lawden,WW}
\begin{gather*}
\frac{\theta_1}{\theta_4}\sim {\rm sn},
\qquad
\frac{\theta_2}{\theta_4}\sim{\rm cn},
\qquad
\sim\frac{\theta_3}{\theta_4}\sim{\rm dn},
\end{gather*}
and describe dynamics of some famous models in the rigid body theories~\cite{landau2,lawden}.
In turn, most popular of them~-- Eulerian top~-- may be quantized, thereby determining the space quantization phenomenon,
so called quantum rotator~\cite{landau}.

These and many other aspects give rise a~problem towards the direct quantization of theta-functions
themselves~\cite{smirnov} if it is possible.
Fortunately this is so indeed and we shall exhibit a~particular case that is completely solvable in the quantum sense:
spectrum of a~quantum Hamiltonian can be found.
The starting point here is the fact that~$\theta$-series posses the nice dif\/ferential properties like elliptic
${\rm sn}$-, ${\rm cn}$-, ${\rm dn}$-functions mentioned above.
In the context above one can even say that any kind of dif\/ferential def\/inition to the $\theta$-functions might be
considered as more appropriate way of their introduction at all.

\section{Definitions and base dynamical systems}

Complete dif\/ferential properties of Jacobian series were described very recently~\cite{brezhnev} and we reproduce them
in a~nutshell.
These four~$\theta$-series are def\/ined as~\cite{abr,WW}
\begin{gather*}
\theta_1(t|\tau) = -{\mathrm{i}} {\mathrm{e}}_{\strut}^{{\frac14}\pi{\mathrm{i}}\tau}
\sideset{}{_k}\sum\limits_{-\infty}^{+\infty} (-1)^k {\mathrm{e}}^{(k^2+k)\pi{\mathrm{i}} \tau} {\mathrm{e}}^{(2k+1)\pi
{\mathrm{i}} t},
\\
\theta_2(t|\tau) = {\mathrm{e}}_{\strut}^{{\frac14}\pi{\mathrm{i}}\tau} \sideset{}{_k}\sum\limits_{-\infty}^{+\infty}
{\mathrm{e}}^{(k^2+k)\pi{\mathrm{i}} \tau} {\mathrm{e}}^{(2k+1)\pi {\mathrm{i}} t},
\\
\theta_3(t|\tau) = \sideset{}{_k}\sum\limits_{-\infty}^{+\infty} {\mathrm{e}}^{k^2\pi{\mathrm{i}} \tau}
{\mathrm{e}}^{2k\pi{\mathrm{i}} t},
\qquad
\theta_4(t|\tau) = \sideset{}{_k}\sum\limits_{-\infty}^{+\infty} (-1)^k {\mathrm{e}}^{k^2\pi{\mathrm{i}} \tau}
{\mathrm{e}}^{2k\pi {\mathrm{i}} t}
\end{gather*}
and satisfy the system of ordinary dif\/ferential equations
\begin{gather}
\begin{split}
&\frac{\partial\theta_2}{\partial t} = \frac{\Dtheta}{\theta_1} \theta_2- \pi \vartheta_2^2
\frac{\theta_3\theta_4}{\theta_1},
\qquad
\frac{\partial\theta_1}{\partial t} =\Dtheta,
\\
& \frac{\partial\theta_3}{\partial t} = \frac{\Dtheta}{\theta_1} \theta_3- \pi \vartheta_3^2
\frac{\theta_2\theta_4}{\theta_1},
\qquad
\frac{\partial\Dtheta}{\partial t} = \frac{\Dtheta{}^2}{\theta_1}-\pi^2\vartheta_3^2
\vartheta_4^2 
\frac{\theta_2^2}{\theta_1}
- \bigg\{4 \eta+\frac{\pi^2}{3}\big(\vartheta_3^4+\vartheta_4^4\big) \bigg\} 
\theta_1,
\\
& \frac{\partial\theta_4}{\partial t} = \frac{\Dtheta}{\theta_1} \theta_4^{}- \pi \vartheta_4^2
\frac{\theta_2\theta_3}{\theta_1},
\end{split}
\label{full}
\end{gather}
where $\Dtheta$ is an additional (f\/ifth) independent function and $(\vartheta, \eta)$ are so
called~$\vartheta$-constants.
In standard theory these `constants' are the values of $\theta(z|\tau)$-functions at $z=0$, i.e., $\vartheta_k=\theta_k(0|\tau)$, but for the dif\/ferential formulation~\eqref{full} such a~meaning need not be
implied; all the $(\eta, \theta_k)$ are allowed to be free external parameters~\cite{brezhnev}.
In the abstract context the dif\/ferential introduction of functions is much more preferable because we may consider these
equations merely as a~vector f\/ield and freely change coordinates on a~manifold where the f\/ield acts.

For a~quantization procedure it is desirable to simplify vector f\/ields into the polynomial ones.
It appears that the sought-for transformation $(\theta_1,\theta_2,\theta_3,\theta_4,\Dtheta)
\rightleftarrows(x,y,z,u,\xi)$ does exist and has the form
\begin{gather*}
\left\{
x=-\pi\vartheta_3\vartheta_4 \frac{\theta_2}{\theta_1},
\;
y=-\pi\vartheta_2\vartheta_4 \frac{\theta_3}{\theta_1},
\;
z=-\pi\vartheta_2\vartheta_3 \frac{\theta_4}{\theta_1},
\;
u={\mathrm{e}}^{\frac{1}{2}\Lambda t^2}_{\mathstrut}
\theta_1,
\;
\xi=\frac{\Dtheta}{\theta_1}+\Lambda t
\right\}
\end{gather*}
under certain value of the parameter $\Lambda(\vartheta,\eta)$.
Let dot above symbol stand for a~$t$-derivative.
Then this substitution brings the~$\theta$-system above into a~very simple polynomial form
\begin{gather}
\label{5D}
{\dot x=y z,
\qquad
\dot y=x z,
\qquad
\dot z=x y^{\mathstrut}}{},
\qquad
\dot\xi=-x^2_{\mathstrut}{},
\qquad
\dot u=\xi u
\end{gather}
which has an embedded structure like `Russian dollies' $(((x,y,z),\xi),u)$.
Moreover it is not dif\/f\/icult to show that the f\/irst three equations in~\eqref{5D} are nothing but the normalized
(parameter-free) model of a~free Euler's rigid body~\cite{landau2,lawden}
\begin{gather*}
\dot X =\left(\frac 1C-\frac1B\right) YZ,
\qquad
\dot Y =\left(\frac 1A-\frac1C\right) XZ,
\qquad
\dot Z =\left(\frac 1B-\frac1A\right) XY
\end{gather*}
with the standard Hamilton and Casimir functions
\begin{gather*}
H=\frac{1}{2}\bigg(\frac{X^2}{A}+\frac{Y^2}{B}+\frac{Z^2}{C} \bigg),
\qquad
K=X^2+Y^2+Z^2
\end{gather*}
and the well-known Poisson bracket
\begin{gather}
\label{lie}
\left[
\begin{matrix}
\smash{\dot X}
\\
\smash{\dot Y}
\\
\smash{\dot Z}
\end{matrix}
\right]= \left[
\begin{matrix}
0 & -Z & Y
\\
Z & 0 & -X
\\
-Y & X & 0
\\
\end{matrix}
\right] \left[
\begin{matrix}
H_X
\\
H_Y
\\
H_Z
\end{matrix}
\right].
\end{gather}
The transformation $(x,y,z)\rightleftarrows(X,Y,Z)$ has a~linear form
\begin{gather*}
X =-\frac{A\sqrt{BC(C-B)}}{\sqrt{(C-B)(A-C)(B-A)}} x,
\qquad
Y =-\frac{B\sqrt{CA(A-C)}}{\sqrt{(C-B)(A-C)(B-A)}} y,
\\
Z =-\frac{C\sqrt{AB(B-A)}}{\sqrt{(C-B)(A-C)(B-A)}} z
\end{gather*}
and holds only for the full asymmetric top parameters $A\ne B\ne C\ne A$.
Quantization of this dynamical system is very well known as the quantum rotator~\cite{landau}.
We thus arrive at a~quantization problem related to a~5D-\emph{extension} of the classical 3D quantized rotator.

As far as we know, no regular algorithm for searching for a~Poisson bracket exists.
Even Euler's top is not an exception since its Lie algebraic $\mathbb{SO}(3)$-bracket is ad~hoc; from the Lie algebra
$\mathfrak{so}(3)$ to the Poisson structure~\eqref{lie}.
On the other hand, it is naturally to expect that knowing the solution in full details one could be able to f\/ind out all
the associated constructions.
Particularly, classical and all the other Poisson structures because they are closely related to integrals of dynamical
systems.
We thus need a~complete set of these functions.

\section{Integrals of motion}\label{3}

Since equations~\eqref{5D} is an odd dimensional system we restrict ourselves in this work to its \hbox{4-di}\-mensional
subsystem
\begin{gather}
\label{sys}
\dot x=y z,
\qquad
\dot y=x z,
\qquad
\dot z=x y^{\mathstrut},
\qquad
\dot\xi=-x^2_{\mathstrut}{}.
\end{gather}
The point here is that there are some serious foundations that even the full 5D system~\eqref{5D} may not be considered
as closed and requires further extension~\cite{brezhnev}.
Along with the elliptic functions and meromorphic elliptic integrals, the logarithmic integral should be involved to the
complete theory~\cite{brezhnev}.
Yet another aspect needs to be kept in mind is that we get in~\eqref{sys} the even dimensional vector f\/ield and may
compare it locally with a~straightened one having (always and at least) the canonical symplectic structure~\cite{olver}.
Actually, this program will be realized in this and next sections.

The functional form of solution to~\eqref{sys} follows from the widely known elliptic solutions of the Euler top
followed by supplementing the meromorphic Legendre integral $\boldsymbol E$ (see~\cite{abr,WW} and appendix for notation
and details):
\begin{gather}
x=-k \alpha\,{\rm sn}(\alpha t+\varepsilon;k),
\qquad
y={\mathrm{i}} \alpha\,{\rm dn}(\alpha t+\varepsilon;k),
\qquad
z={\mathrm{i}} k \alpha\,{\rm cn}(\alpha t+\varepsilon;k),
\nonumber
\\
\xi=-k^2\alpha^2\int^{ t} {\rm sn}^2(\alpha s+\varepsilon;k) ds=K+\alpha \mathtt{Z}(\alpha
t+\varepsilon;k)-\alpha^2 t,
\label{mero}
\end{gather}
where $(k,\alpha,K;\varepsilon)$ are free constants.
However, we are about to get some invariant constructions (Poisson bracket etc) and therefore one should transform this
solution into the integrals of motion form.

Let $(I,J,K)$ denote the three independent functions of $(x,y,z,\xi)$ being constant on solutions of~\eqref{sys} and let
${\mathcal{N}}$ be a~function of the same variable set evolving linearly in time.
Then the locally equivalent form to the system under consideration is
\begin{gather}
\label{pi}
\dot{\mathcal{N}}=1,
\qquad
\dot I=0,
\qquad
\dot J=0,
\qquad
\dot K=0.
\end{gather}
We shall informally name the $({\mathcal{N}};I,J,K)$-set as `straightened' variables $(\pi^1,\ldots,\pi^4)$.
Clearly, the pair $(I,J)$ is an equivalent of $(\alpha,k)$ and ${\mathcal{N}}$ is def\/ined up to a~function of
$(I,J,K)$, that is the constant~$\varepsilon$.
To put the scheme dif\/ferently we are seeking for a~formal point transformation
$(x,y,z,\xi)\rightleftarrows(\pi^1,\ldots,\pi^4)$ between the original and straightened variable set in both directions.
Then some routine computations with elliptic integrals yield the sought-for change:
\begin{gather*}
x =-I\,{\rm sn}\left(J{\mathcal{N}}; {\frac IJ}\right),
\qquad
y= {\mathrm{i}} J\, {\rm dn}\left(J{\mathcal{N}}; {\frac IJ}\right), \qquad
z= {\mathrm{i}} I\,{\rm cn}\left(J{\mathcal{N}}; {\frac IJ}\right),
\\
\xi=K+J \mathtt{Z}\left(J{\mathcal{N}}; {\frac IJ}\right)-J^2{\mathcal{N}},\\
{\mathcal{N}}=\frac{1}{\sqrt{x^2-y^2}} {\boldsymbol F}\left(\frac{x}{\sqrt{x^2-z^2}};
\frac{\sqrt{x^2-z^2}}{\sqrt{x^2-y^2}}\right),
\qquad I^2=x^2-z^2,
\qquad J^2 =x^2-y^2,\\
 K =\xi+\sqrt{x^2-y^2}
\left\{{\boldsymbol F}\left(\frac{x}{\sqrt{x^2-z^2}}; \frac{\sqrt{x^2-z^2}}{\sqrt{x^2-y^2}}\right)
 -{\boldsymbol E}\left(\frac{x}{\sqrt{x^2-z^2}}; \frac{\sqrt{x^2-z^2}}{\sqrt{x^2-y^2}}\right)\right\}.
\end{gather*}
Hence the system~\eqref{sys} has only two independent single-valued (rational) integrals~$I$ and~$J$ and the Hamilton
function~$H$ may be chosen only as their combination: $H=H(I,J)$.
Dif\/ferential computations with all the functions (${\rm sn}$, ${\rm cn}$, ${\rm dn}$, $\mathtt{Z}$,
$\boldsymbol F$, $\boldsymbol E$) have been completely developed; see the well-known Jacobi calculus~\cite{abr, WW}
accompanied with formulas~\eqref{FE}--\eqref{Z}.

\section{Poisson bracket(s)}

The Hamilton function and Poisson bracket~$\Omega$ appear always in a~pair and the former is subjected to our free assignment.
On the other hand, every vector f\/ield is locally equivalent to the simplest one, like~\eqref{pi}, and its Poisson
structure may be assigned among the constant symplectic ones.
For comparison with the standard Lie algebraic case~\eqref{lie} let us consider the linear in $I^2$, $J^2$ case $a I^2+b J^2$:
\begin{gather*}
H=a \big(x^2-z^2\big)+b \big(x^2-y^2\big)
\end{gather*}
with some constants $a$, $b$.

\subsection*{The method}

The scheme of f\/inding the bracket is based on the fact that under f\/ixed dimension any bracket for any hamiltonian system
is locally equivalent to any other bracket.
Hence we may take as an ansatz some `known and good, straightened' bracket ${\mathsf{K}}^\mathit{jk}(\pi)$
for known `simple' vector f\/ield $\dot \pi^k=v^k(\pi)$ and seek for its transformation to the initial f\/ield coordinates~$(x^k)$.
The $x$-coordina\-tes~$\Omega^\mathit{jk}(x)$ of the bracket tensor is then recalculated by the standard transformation
law
\begin{gather}
\label{OK}
\Omega^{\mathit{jk}}(x)=\Partial{x^j}{\pi^n} \Partial{x^k}{\pi^m}   {\mathsf{K}}^{\mathit{nm}}(\pi).
\end{gather}
It should be noted here that ansatz for ${\mathsf{K}}(\pi)$ may contain some freedom which is not restricted
by the constant parameters (canonical constant bracket ${\mathsf{K}}$) because the only condition for
a~bracket, apart from antisymmetry, is the Jacobi identity.
In terms of inverse matrix-tensor $\boldsymbol{\mathsf{K}}={\mathsf{K}}^{-1}$ the condition turns
out to be a~Maxwell (linear) equations f\/irst half $d \boldsymbol{\mathsf{K}}=0$ which, clearly, has
inf\/initely many solutions.

To continue further quantization we have to produce at least one nontrivial solution.
Some experimental computations show that there exists an unusual bracket
\begin{gather}
\label{so3}
\left[
\begin{matrix}
\smash{\dot x}
\\
\smash{\dot y}
\\
\smash{\dot z}
\\
\smash{\dot \xi}
\end{matrix}
\right]= { \underbrace{\left[
\begin{matrix}
0 & 0 & y\hspace{0.5em} \smash{\raise-2.85em\hbox{\rule{0.3pt}{3.6em}}}\hspace{-0.5em} & -x
\\
0 & 0 & x & -y
\\
\smash{\raise-0.9ex\hbox{\rule{5.45em}{0.3pt}}}\hspace{-5.45em}{-}y & -x & 0 & 0
\\
x & y & 0 & 0
\end{matrix}
\right]} } \left[
\begin{matrix}
H_x
\\
H_y
\\
H_z
\\
H_\xi
\end{matrix}
\right],
\qquad
H=\frac{1}{2} \big(z^2-x^2\big)
\\
\hspace{7.3em}\Omega
\nonumber
\end{gather}
and its determinant is not a~zero in the generic case: $\det\Omega=(x^2-y^2)^2$.
Interestingly enough, the upper-left $(3{\times}3)$-submatrix of~$\Omega$ (boxed) is also a~bracket and provides yet
another, rather nonstandard, bracket for Euler's top; it is related to the $\mathfrak e(1,1)$-algebra.
For reference we display also the commuting vector f\/ield corresponding to the integral $I^2=y^2-x^2$:
\begin{gather*}
\dot x=0,
\qquad
\dot y=0,
\qquad
\dot z=0,
\qquad
\dot \xi= 2 \big(y^2-x^2\big),
\qquad
\{H,I\}=0.
\end{gather*}

\begin{remark}\label{remark1}
Informally, we may conclude that knowing the bracket tensor is an equivalent to knowing the dif\/ferential computation
rules for the functions realizing complete system of integrals for the dynamics.
Appearance/introduction of a~specif\/ic form to the functions may be thought of as a~reduction of the full system
dimension to a~smaller number of functions; the absence of integrals means thus just a~generic~-- nonintegrable or not
yet integrated~-- case.
If we have dif\/ferential function calculus in one direction, say, $(x)\stackrel{\partial}{\dashrightarrow}(\pi)$, then
calculus in a~reverse direction $(\pi)\stackrel{\partial}{\dashrightarrow}(x)$ is automatically induced.
\end{remark}

\begin{example}
Consider the planar vector f\/ields
\begin{gather*}
\dot x=A(x,y),
\qquad
\dot y=B(x,y),
\end{gather*}
where~$A$ and~$B$ are some polynomials in $(x,y)$ over $\mathbb{C}$.
Recently there was developed and implemented an ef\/fective algorithm~\cite{weil} for computing the rational integral
$H=R(x,y)$ to these equations.
In this connection P.~Olver put a~question (October 2014): what about Hamiltonian structure for these systems? As we
have seen from scheme~\eqref{OK}, this question has a~constructive answer.

First, let us straighten the f\/ield under question.
One of the $(\pi^1,\pi^2)=({\mathcal{N}},H)$-variables we've already had; this is the
rational~$H$-Hamiltonian $R(x,y)$ coming from the algorithm mentioned above.
The second variable~-- momentum conjugated to~$H$~-- is derived from equations determining trajectory of the system.
Indeed,
\begin{gather*}
t=\int \frac{dx}{A(x,y)},
\qquad
R(x,y)=\mathrm{const}
\end{gather*}
and, hence, variable ${\mathcal{N}}$ is an integral
\begin{gather*}
{\mathcal{N}}=\int^{ x} \frac{dz}{A(z,w)}\FED\aleph(x;R(x,y))
\end{gather*}
and this integral is an Abelian one with respect to algebraic irrationality
\begin{gather}
\label{curve}
R(z,w)=\mathrm{const}.
\end{gather}
Here $\aleph(x;\mathrm{const})$ is a~function representation to the integral as function of the~$x$-variable and
a~constant being a~modulus of the algebraic curve~\eqref{curve}.
Dif\/ferential computations for this Abelian integral should be developed independently; including, which is an important
and nontrivial point (see Remark~\ref{remark1}), dif\/ferentiation with respect to `const'
of the curve~\eqref{curve}.
Since $\aleph$ is an Abelian integral its dif\/ferential calculus will possibly involve Abelian integrals of other kinds
as well as rational functions of the pair $(x,y)$~\cite{baker}.
Anyway, development of the calculus is a~technical and solvable problem because derivative of the Abelian integral in
a~parameter is again an integral belonging to the same algebraic curve.

Now, we assign the function $R(x,y)$ as a~Hamiltonian $H=R(x,y)$ and take the simplest canonical local
$(2{\times}2)$-bracket ${\mathsf{K}}=\left(
\begin{smallmatrix}
\phantom{-}0&1
\\
-1&0
\end{smallmatrix}
\right)$ for the straightened vector f\/ield
\begin{gather*}
\dot{\mathcal{N}}=1,
\qquad
\dot H=0.
\end{gather*}
Then the sought-for $(x,y)$-coordinates $\Omega(x,y)$ of the bracket is calculated explicitly by~\eqref{OK}.
Suf\/f\/ice it to display only the $\Omega^{12}$-component:
\begin{gather}
\label{comp}
\left(
\begin{matrix}
\dot x
\\
\dot y
\end{matrix}
\right)=\Omega \nabla R,
\qquad
\big(\Omega^{12}\big)^{-1}= \nabla_{y} R \cdot\nabla_{x} \aleph- \nabla_{x} R \cdot\nabla_{y} \aleph.
\end{gather}
Rationality or the global single-valuedness of the found bracket tensor is established (if any) on the base of
dif\/ferentiation rules for the Abelian integrals that arise.
Then it remains only to reassign the Hamiltonian $H\to f(H)$ for a~simplif\/ication of the f\/inal answer.

Let us consider a~further specif\/ic subexample: the vector f\/ield
\begin{gather*}
\dot x=y,
\qquad
\dot y=6 x^2
\end{gather*}
def\/ining the equianharmonic elliptic Weierstrass function $x=\wp(t;0,a)$ and its derivative $y=\wp'(t;0,a)$~\cite{WW}.
According to the scheme above, we have
\begin{gather*}
H=\frac12\big(y^2-4 x^3\big),
\qquad
{\mathcal{N}}=\int^{ x} \frac{ds}{\sqrt{4 s^3-\mathrm{const}}}
\quad
\Rightarrow
\quad
{\mathcal{N}}= \mathfrak{A}\big(x;0,4 x^3-y^2\big),
\end{gather*}
where $\mathfrak{A}(x;g_2^{},g_3^{})$ is a~notation for a~holomorphic integral in the Weierstrass elliptic
theory~\cite{WW}.
Although dif\/ferential calculus with Weierstrassian integrals is also absent in standard literature, it can be developed
in the same manner as Legendrian formulas~\eqref{FE}, \eqref{PI}.
Computation~\eqref{comp} with rational function $R(x,y)$ and transcendental one $\mathfrak{A}(x;0,4 x^3-y^2)$ yields:
$\Omega^{12}=1$.
This (global) unity corresponds exactly to the standard canonical bracket $\Omega=\left(
\begin{smallmatrix}
\phantom{-}0&1
\\
-1&0
\end{smallmatrix}
\right)$.
It should be emphasized here that the seed `straightened' bracket ${\mathsf{K}}$ is not well-def\/ined but is
just a~`local unity' since ${\mathcal{N}}=\mathfrak{A}(x;0,4 x^3-y^2)$ is a~multivalued function.
Meanwhile the $(x,y)$-coordinates of the bracket are the global well-def\/ined `object unity'.
Clearly, the $(2{\times}2)$-case under consideration has an additional convenience.
Reassignment of the Hamilton function exhausts all the possible brackets and there is no need to investigate nonconstant
`straightened' brackets ${\mathsf{K}}({\mathcal{N}},H)$.
\end{example}

\begin{example}
In a~slightly dif\/ferent but equivalent manner the Hamilton~$\tau$-dynamics was found in work~\cite{br3} for the elliptic
$\theta(0|\tau)$-constants; the very $\vartheta$, $\eta$-coef\/f\/icients of system~\eqref{full}.
\end{example}

\begin{example}[exercise]
Extend the method into odd dimensional cases and reproduce the standard Poisson
$\mathbb{SO}(3)$-description~\eqref{lie}.
\end{example}

\section{Quantization: operators and spectrum}

In this section we propose a~complete canonical quantization procedure for the model~\eqref{so3} described above.
As a~byproduct, we get also some nonstandard quantization for the classical Euler equations~\eqref{lie}.

As usual, we should, according to the correspondence principle
\begin{gather*}
\{x,y\}\dashrightarrow[\what x,\skew1\what y],
\end{gather*}
f\/ind representations for the phase observable operators $(x,y,z,\xi)\dashrightarrow (\what x,\skew1\what y,\what z,\skew1\what \xi)$, where
bracket $\{x,y\}$ is given by the~$\Omega$-matrix~\eqref{so3}.
Taking into account that Euler's $(3{\times}3)$-subsystem~\eqref{lie} is realized by the 1st order dif\/ferential
operators with linear coef\/f\/icients (the classical momentum operators~\cite{landau}), we can try the same structure
ansatz for our $(4{\times}4)$-extension
\begin{gather*}
(\text{phase variable operator}) = (\text{linear function})\times \partial_{\text{variable}}+(\text{linear function}).
\end{gather*}
Note, the standard generators $\partial_x$, $\partial_y$, $x\partial_y+y\partial_x$ of the aforementioned
$\mathfrak{e}(1,1)$-algebra are also nicely f\/it this ansatz.
Such a~solution~-- combination of dilatations, rotations, and $\partial$-operators~-- does exist and we f\/ind
\begin{gather}
\label{repr}
\what x\DEF x,
\qquad
\skew1\what y\DEF y,
\qquad
\what z\DEF -x \partial_y-y \partial_x,
\qquad
\skew1\what \xi \DEF x \partial_x+y \partial_y
\end{gather}
(f\/irst pointed out to the author by P.~Kazinsky).
Commuting relations
\begin{gather*}
[\what x, \what z]=\skew1\what y,
\qquad
[\skew1\what y,\what z]= \what x,
\qquad
[\skew1\what \xi,\what x]=\what x,
\qquad
[\skew1\what \xi, \skew1\what y]=\skew1\what y
\end{gather*}
are the same as the bracket ones for base variables $(x,y,z,\xi)$, i.e.,~\eqref{so3}:
\begin{gather*}
\{x,z\}=y,
\qquad
\{y,z\}=x,
\qquad
\{\xi,x\}=x,
\qquad
\{\xi,y\}=y
\end{gather*}
(all the absent pairs are zeroes).
For simplicity we omit here the physical multipliers like ${\mathrm{i}} \hbar$ in front of commutators.
From~\eqref{repr} it follows immediately that quantum Hamiltonian is represented by the operator
\begin{gather*}
\widehat{\mathscr{H}}\DEF\frac12 \big(\what z^2-\what x^2\big)= \frac12 \big\{(x \partial_y+y \partial_x)^2-x^2\big\}
\end{gather*}
and we readily check that quantum equations of motion (Heisenberg's dynamics) also holds with respect to the canonical
Weyl ordering
\begin{gather*}
\what{\dot x}=\frac12 (\skew1\what y\what z+\what z\skew1\what y) =[ \what x, \widehat{\mathscr{H}}],
\\
\skew1\what{\dot y} =\frac12 (\what x\what z+\what z\what x)=[ \skew1\what y, \widehat{\mathscr{H}}],
\\
\skew2\what{\dot z}=\frac12 (\what x\skew1\what y+\skew1\what y\what x) =[ \what z, \widehat{\mathscr{H}}],
\\
\skew2\what{\dot \xi} =-\frac12 (\what x\what x+\what x\what x) =[ \skew1\what \xi, \widehat{\mathscr{H}}].
\end{gather*}
Availability of a~(complex) rotation operator $x \partial_y+y \partial_x$ in $\widehat{\mathscr{H}}$ prompts us to
make a~variable change to the radius-angle $(r,\gamma)$-variables
\begin{gather*}
x=2 r\cos\big(\tfrac12\gamma\big),
\qquad
y=2 {\mathrm{i}} r\sin\big(\tfrac12\gamma\big),
\end{gather*}
whereupon we get $\what z=-2 \partial_\gamma$ and the following equivalent of the eigen-value problem
\begin{gather*}
\widehat{\mathscr{H}}\Psi=E \Psi
\quad
\Rightarrow
\quad
\Psi_{\gamma\gamma}+ \frac{1}{2} r^2\cos(\gamma)\, \Psi=-\frac12 \big(E+r^2\big) \Psi.
\end{gather*}
Variable~$r$ (commuting observable~$I$) has been separated, i.e., became a~parameter, and we may f\/inally normalize the
spectral problem into the Schr\"odinger equation with a~trigonometric cos-type potential
\begin{gather*}
\Psi''=(A\cos\boldsymbol\gamma-E)\Psi
\end{gather*}
and two extra parameters~$E$ and~$A$.
This is the very well-known equation~-- Mathieu equa\-tion~\cite{abr, WW}~-- and excellent book on its theory has hitherto
remained the MacLachlan treatise~\cite{mac}.

Until now, we were dealing with integrable but complex equations.
Their general integrability is related essentially to the complex phase space $\mathbb{C}^4$ but for physical motivation
we need a~real reduction of the space as long as the reduction was a~non-contradictory one; see also some arguments on
reality and complexif\/ication at the end of \S~1 in~\cite{smirnov}.
A~rational question at this point is: Whether this model is realistic? In other words, is it possible to `adjust' our
formal (yet complexif\/ied) quantities in system~\eqref{sys} to be real physical observables and their operators to be
(real) self-adjoining ones? The question has a~positive answer.
Indeed, we want to have real~$H$ and, hence, $x^2,z^2,\gamma, r^2\in \mathbb{R}$ under $y^2-x^2\ne 0$
($\det\Omega\ne0$).
These requirements can be satisf\/ied if we put $x,\xi\in\mathbb{R}$, $y\in{\mathrm{i}}\mathbb{R}$,
$z\in{\mathrm{i}}\mathbb{R}$ and $\gamma,r\in\mathbb{R}$.
So, changing $z\to{\mathrm{i}} z$, $y\to {\mathrm{i}} y$ we keep the system~\eqref{sys} to be pure real and Hamiltonian
\begin{gather*}
\widehat{\mathscr{H}} \dashrightarrow \frac12\big(\what z^2+\what x^2\big)
\end{gather*}
then becomes not only the real operator but also a~lower bounded one.
It immediately follows that background state and quantum mechanics do exist and therefore this Hamiltonian may be `not
a~bad' candidate for the realistic one with a~physical spectrum.

Mention should be made of relations between proposed scheme and some closely related models.
On the one hand, the Mathieu equation can be treated as a~quantum physical \hbox{1-d}i\-men\-sional pendulum and its
limit $A\cos \gamma =A \big(1-\frac12 \gamma^2+\cdots\big)$ reproduces locally a~quadratic oscillator with an inf\/inite
discrete spectrum.
Parameter~$A$ may play here a~role of the potential well depth.
The model~\eqref{sys} does thus inherit properties of these limit cases (a dimension reduction $4 {\dashrightarrow} 2$
and approximation $\cos \gamma {\dashrightarrow} \gamma^2$) and, on the other hand, is an alternative to the standard
$\mathbb{SO}(3)$-quantization of a~rotator~\cite{landau}.
It is interesting to observe that quantum statements in both this quantization scheme and proposed new one are
determined by the complete set of compatible/commuting observables which are classically evolving as \emph{meromorphic
functions} of~$t$; not as meromorphic/logarithmic elliptic integrals.
Recall, meromorphic (elliptic) functions are the single-valued (well-def\/ined) functions on torus, whereas integrals, in general, are not.
Indeed, the standard momentum square $\boldsymbol L^2$ and $L_z$ for the rigid Euler's body are combinations of
${\rm sn}$-, ${\rm cn}$-, and ${\rm dn}$-functions~\cite{landau}.
In the model under consideration the variable $\xi$, which is a~meromorphic elliptic integral~\eqref{mero}, does not get
to the commuting pair of observables and all the other phase variables are also the meromorphic functions on a~torus.
We observe in passing that integral structure of the classical solutions to this new model is much simpler than the
standard one.
One of the Euler classical position $(\psi,\theta,\varphi)$-angles of a~body~-- the angle~$\varphi(t)$~-- is a~cumbersome
combination of meromorphic and logarithmic elliptic integrals; see~\cite[Chapter~VI, \S~37, equations~(37.17)--(37.20)]{landau2}
and~\cite[p.~139]{lawden}.
Meanwhile the only variable of an integral type in the model~\eqref{sys}~-- variable~$\xi$~-- is just a~simple
$\mathtt{Z}$-function or, which is the same, some $\Dtheta/\theta$-combination.
One may be noted in this connection that complete quantization procedure of theta functions will perhaps inevitably
involve, the holomorphic integrals excepted, the full set of Abelian objects on Abelian variety: meromorphic functions,
meromorphic and logarithmic integrals, and, which is the most nontrivial point, the~$\theta$ itself as integral of an
Abelian meromorphic integral.
Such a~nature of the theta-function is explained in~\cite{brezhnev}.

\begin{figure}[t] \centering \includegraphics[width=10.8cm]{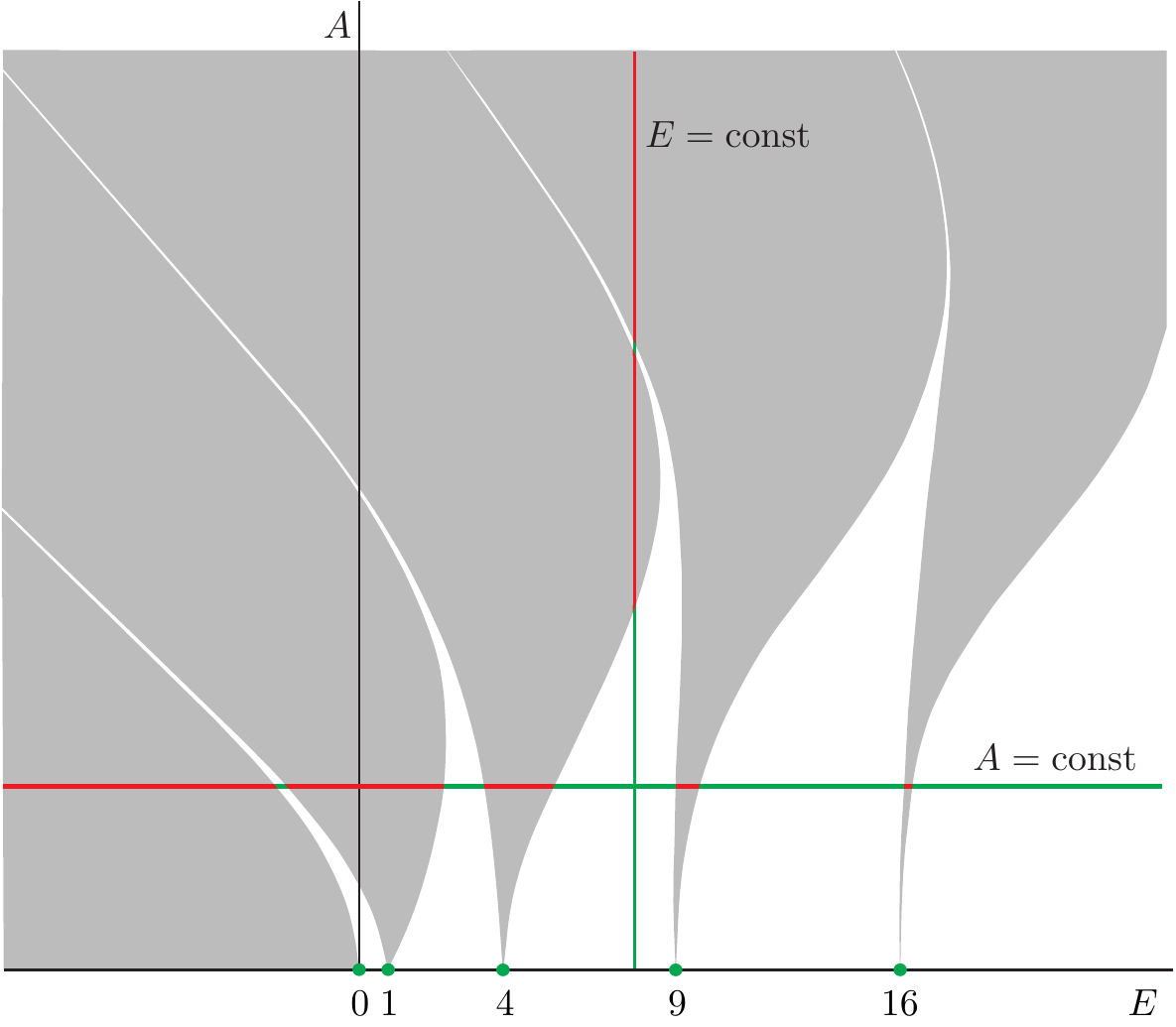}
\caption{The gap structure spectrum for the Mathieu equation $\Psi''=(A\cos\boldsymbol\gamma-E) \Psi$.
Shaded domains are the spectral lacunae (`forbidden' zones).}\label{fig}
\end{figure}

Finally, Fig.~\ref{fig} exhibits the typical spectra structure for the Mathieu equation~\cite{nayf} in both the quantum
numbers~$E$ and~$A$.
The number of lacunae (colored in red) is inf\/inite and bold green dots on the horizontal~$E$-axis is the sequence $0^2$,
$1^2$, $2^2$, \ldots\ignorespaces.
It is a~gap boundary set of the limit case $A\to0$ following from the fact that the zero/constant potential has pure
$2\pi$-periodic solutions when spectral parameter is an integer perfect square $E=n^2$.
Exhaustive theory, numerous applications, and wealth of numeric information to the Fig.~\ref{fig} can be found
in~\cite{ince,mac,yakub}.
Note that number of gaps in~$A$-direction is also inf\/inite as in the~$E$-axis; for proof see~\cite[Chapter~7, \S~3.5]{yakub}.

\appendix

\section{Appendix.
Dif\/ferential calculus of Legendre integrals}

The standard Legendrian elliptic integrals~\cite{abr} are the holomorphic and meromorphic integrals
\begin{gather*}
\boldsymbol F(x;k)\DEF\int_0^{ x} \frac{ds}{\sqrt{(1-s^2)(1-k^2s^2)}},
\qquad
\boldsymbol E(x;k)\DEF\int_0^{ x} \frac{\big(1-k^2s^2\big) ds}{\sqrt{(1-s^2)(1-k^2s^2)}}
\end{gather*}
(1st and 2nd kind respectively) and the logarithmic integral (3rd kind)
\begin{gather*}
{\boldsymbol\Pi_\alpha(x;k)\DEF \int_0^{ x} \frac{1}{1-\alpha s^2} \frac{ ds}{\sqrt{(1-s^2)(1-k^2s^2)}}}.
\end{gather*}
The latter has a~parameter~$\alpha$.

The triple $(\boldsymbol F, \boldsymbol E, \boldsymbol\Pi_\alpha)$ forms a~linear vector space and the theory above
requires complete set of its partial dif\/ferentiations $\partial_x$, $\partial_k$, and $\partial_\alpha$.
Somewhat surprising facet is the fact that no self-contained dif\/ferential calculus of these objects with respect to
\emph{all} the arguments $(x, k, \alpha)$ seems to have hitherto been tabulated in handbooks; see, e.g.,~\cite{abr,bateman,tannery}.
Bearing in mind a~remark about logarithmic integral at the beginning of Section~\ref{3}, we make up this def\/iciency.

Denote $y^2\DEF(1-x^2)(1-k^2x^2)$.
Then one can show that the three objects $(\boldsymbol F, \boldsymbol E, \boldsymbol\Pi_\alpha)$ have the following
\emph{closed} rules of dif\/ferential computations
\begin{gather}
\begin{split}
& \Partial{\boldsymbol F}{x} =\frac{1}{y},
\qquad
 \Partial{\boldsymbol F}{k} =-\frac{\boldsymbol F}{k}-\frac{\boldsymbol E}{k (k^2-1)}+
\frac{k}{k^2-1} \frac{x y}{1-k^2x^2},
\\
& \Partial{\boldsymbol E}{x} =\big(1-k^2x^2\big) \frac1y,
\qquad
\Partial{\boldsymbol E}{k} =-\frac{\boldsymbol F}{k}+\frac{\boldsymbol E}{k\mathstrut},
\end{split}
\label{FE}
\\
\begin{split}
&\Partial{\boldsymbol\Pi_\alpha}{x} =\frac{1}{1-\alpha x^2} \frac1y,
\\
&\Partial{\boldsymbol\Pi_\alpha}{k} =\frac{-k}{(k^2-\alpha)(k^2-1)} \left(\boldsymbol E+\big(k^2-1\big)
\boldsymbol\Pi_\alpha-\frac{k^2 x y}{1-k^2x^2} \right),
\\
&\Partial{\boldsymbol\Pi_\alpha}{\alpha} = \frac{1}{2 \alpha (\alpha-1)(k^2-\alpha)}\left(\big(k^2-\alpha\big)\boldsymbol
F+\alpha \boldsymbol E-(k^2-\alpha^2) \boldsymbol\Pi_\alpha- \frac{\alpha^2  x y}{1-\alpha x^2}\right).
\end{split}
\label{PI}
\end{gather}
Note that these equations are compatible; the cross derivatives like
\begin{gather*}
\Partial{}{x}\Partial{\boldsymbol E}{k}=\Partial{}{k}\Partial{\boldsymbol E}{x},
\qquad
\Partial{}{\alpha}\Partial{\boldsymbol\Pi_\alpha}{k}= \Partial{}{k}\Partial{\boldsymbol\Pi_\alpha}{\alpha},
\qquad
\ldots
\end{gather*}
turn out to be identities.
If we introduce also a~Jacobi analog of Weierstrassian~$\zeta$-function, i.e., $\mathfrak{u}$-rep\-re\-sen\-ta\-tion
\begin{gather*}
\mathtt{Z}(\mathfrak{u};k)\DEF\boldsymbol E\big({\rm sn}(\mathfrak{u};k);k\big)
\end{gather*}
of the $\boldsymbol E$-integral, then it is dif\/ferentiated as follows
\begin{gather}
\label{Z}
\Partial{\mathtt{Z}}{\mathfrak{u}}=1-k^2\,{\rm sn}^2,
\qquad
\Partial{\mathtt{Z}}{k}=\frac{k}{k^2-1}\big(\mathtt{Z}\,{\rm cn}^2-
\big(k^2-1\big)\,\mathfrak{u}\,{\rm sn}^2-{\rm sn}\,{\rm cn}\,{\rm dn}\big).
\end{gather}
If required, partial~$k$-derivatives of functions ${\rm sn}$, ${\rm cn}$, ${\rm dn}(\mathfrak{u};k)$ are calculated
by formulae which are known~\cite[pp.~82--83]{lawden}; we do not display them here.

\subsection*{Acknowledgements}

The author would like to thank Dima Kaparulin and Peter Kazinsky for stimulating discussions and my special thanks are
addressed to S.~Lyakhovich and A.~Sharapov for valuable consultations.
Also, much gratitude is extended to the anonymous referee for helpful suggestions and construc\-tive criticism, which
resulted in considerable improvement of the f\/inal text.
The study was supported by the Tomsk State University Academic D.~Mendeleev Fund Program.

\pdfbookmark[1]{References}{ref}
\LastPageEnding


\begin{thebibliography}{99}
\footnotesize \itemsep=0pt

\bibitem{abr}
Abramowitz M., Stegun I.A., Handbook of mathematical functions with formulas,
  graphs, and mathematical tables, \textit{National Bureau of Standards Applied
  Mathematics Series}, Vol.~55, U.S.~Government Printing Of\/f\/ice, Washington,
  D.C., 1964.

\bibitem{baker}
Baker H.F., Abelian functions. Abel's theorem and the allied theory of theta
  functions,, Cambridge Mathematical Library, Cambridge University Press,
  Cambridge, 1995.

\bibitem{weil}
Bostan A., Ch\`eze G., Cluzeau T., Weil J.-A., Ef\/f\/icient algorithms for
  computing rational f\/irst integrals and Darboux polynomials of planar
  polynomial vector f\/ields, \href{http://arxiv.org/abs/1310.2778}{arXiv:1310.2778}.

\bibitem{brezhnev2}
Brezhnev Yu.V., What does integrability of f\/inite-gap or soliton potentials
  mean?, \href{http://dx.doi.org/10.1098/rsta.2007.2056}{\textit{Philos. Trans.~R. Soc. Lond. Ser.~A}} \textbf{366} (2008),
  923--945, \href{http://arxiv.org/abs/nlin.SI/0505003}{nlin.SI/0505003}.

\bibitem{brezhnev}
Brezhnev Yu.V., Non-canonical extension of {$\theta$}-functions and modular
  integrability of {$\vartheta$}-constants, \href{http://dx.doi.org/10.1017/S0308210512001023}{\textit{Proc. Roy. Soc. Edinburgh
  Sect.~A}} \textbf{143} (2013), 689--738, \href{http://arxiv.org/abs/1011.1643}{arXiv:1011.1643}.

\bibitem{br3}
Brezhnev Yu.V., Lyakhovich S.L., Sharapov A.A., Dynamical systems def\/ining
  {J}acobi's {$\vartheta$}-constants, \href{http://dx.doi.org/10.1063/1.3662961}{\textit{J.~Math. Phys.}} \textbf{52}
  (2011), 112704, 21~pages, \href{http://arxiv.org/abs/1012.1429}{arXiv:1012.1429}.

\bibitem{bateman}
Erd{\'e}lyi A., Magnus W., Oberhettinger F., Tricomi F.G., Higher
  transcendental functions, {V}ol.~{III}, McGraw-Hill Book Company, Inc., New
  York~-- Toronto~-- London, 1955.

\bibitem{ince}
Ince E.L., Researches into the characteristic numbers of the Mathieu equation,
  \href{http://dx.doi.org/10.1017/S0370164600021866}{\textit{Proc. Roy. Soc. Edinburgh}} \textbf{46} (1927), 20--29.

\bibitem{landau2}
Landau L.D., Lifshitz E.M., Mechanics, Pergamon Press, Oxford~-- London~-- New
  York~-- Paris, 1960.

\bibitem{landau}
Landau L.D., Lifshitz E.M., Quantum mechanics: non-relativistic theory,
  Pergamon Press, London, 1981.

\bibitem{lawden}
Lawden D.F., Elliptic functions and applications, \href{http://dx.doi.org/10.1007/978-1-4757-3980-0}{\textit{Applied Mathematical
  Sciences}}, Vol.~80, Springer-Verlag, New York, 1989.

\bibitem{mac}
McLachlan N.W., Theory and application of {M}athieu functions, Clarendon Press,
  Oxford, 1947.

\bibitem{nayf}
Nayfeh A.H., Introduction to perturbation techniques, \textit{A~Wiley-Interscience
  Publication}, John Wiley \& Sons, New York, 1981.

\bibitem{olver}
Olver P.J., Applications of {L}ie groups to dif\/ferential equations,
  \href{http://dx.doi.org/10.1007/978-1-4684-0274-2}{\textit{Graduate Texts in Mathematics}}, Vol.~107, Springer-Verlag, New York,
  1986.

\bibitem{smirnov}
Smirnov F.A.,
What are we quantizing in integrable f\/ield theory?,
  \textit{St.~Petersburg Math.~J.} \textbf{6} (1995), 417--428, \href{http://arxiv.org/abs/hep-th/9307097}{hep-th/9307097}.

\bibitem{tannery}
Tannery J., Molk J., \'El\'ements de la th\'eorie des fonctions elliptiques,
  Vol.~IV, Gauthier--Villars, Paris, 1902.

\bibitem{vanhaecke}
Vanhaecke P., Algebraic integrability: a survey, \href{http://dx.doi.org/10.1098/rsta.2007.2066}{\textit{Philos. Trans.~R. Soc.
  Lond. Ser.~A}} \textbf{366} (2008), 1203--1224.

\bibitem{WW}
Whittaker E.T., Watson G.N., A course of modern analysis. An introduction to
  the general theory of inf\/inite processes and of analytic functions; with an
  account of the principal transcendental functions, \href{http://dx.doi.org/10.1017/CBO9780511608759}{\textit{Cambridge Mathematical
  Library}, Cambridge University Press}, Cambridge, 1996.

\bibitem{yakub}
Yakubovich V.A., Starzhinskij V.M., Linear dif\/ferential equations with periodic
  coef\/f\/icients, Vols.~1,~2, John Wiley \& Sons, New York~-- Toronto, 1975.

\end{thebibliography}
\end{document}